\documentclass[preprint,12pt]{elsarticle}

\usepackage{amssymb}
\begin{document}

\begin{frontmatter}



\title{Variable Stars in the Globular Cluster NGC 4590 (M68)}


\author{Devesh Path Sariya$^a$$^b$\footnote{E-mail:  devesh@aries.res.in~(Devesh P. Sariya); sneh@aries.res.in~(Sneh Lata) and rkant@aries.res.in (R. K. S. Yadav)}~\footnote{Tel: 0091 05942 233734; Fax No: 0091 05942 233439}, Sneh Lata$^a$, R. K. S. Yadav$^a$}

\address{$^{a}$Aryabhatta Research Institute of Observational Sciences, Manora Peak Nainital 263 002, India.\\
$^{b}$ School of studies in Physics \& Astrophysics, Pt Ravishankar Shukla University, Raipur 492010, India.}
\begin{abstract}

We present results of time series photometry to search for variable 
stars in the field of metal-poor globular cluster NGC 4590 (M68). 
Periods have been revised for 40 known variables and no significant changes 
were found. A considerable change in Blazhko effect for V25 has been detected.
Among nine newly discovered variable candidates, 5 stars are of RRc Bailey 
type variables while 4 stars are unclassified. The variable stars V10, V21, V50 
and V51 are found to be cluster members based on the radial velocity data taken 
from literature.
\end{abstract}
%\end{enumerate}

\begin{keyword}
Globular clusters: stars: variables: RR Lyrae


\end{keyword}

\end{frontmatter}

% \linenumbers

\section{Introduction}
\label{Intro}
Galactic globular clusters are among the oldest population of the Universe.
Their studies are of particular interest to know about the evolution history of the Milky Way.
In a globular cluster, most commonly found variables are
RR Lyrae stars (``cluster-type'' variables).
RR Lyrae being short period variables are easy to detect by means of observational data of few hours.
They have long been used as standard candles, making the distance estimation
of a cluster possible. Studying their pulsational properties enables
the astronomers to probe the internal structure of the stars. 
They are particularly important to study the evolution of low mass stars.
RR Lyrae also play important role in studying the horizontal branch (HB) 
morphology of the globular cluster (Newsham 2007).

NGC 4590 is a metal-poor, moderate distant globular cluster with low reddening.
The fundamental parameters of the cluster are listed in Table~\ref{par}. It is a 
variable-rich Oosterhoff type II (OoII)
globular cluster and has long been an interesting object for the variable stars investigations. 
Shapley (1920) carried out the first investigation of variable stars in the cluster 
and found 27 RR Lyrae type variables. Rosino and Pietra (1954) found 3 additional RR Lyrae variables
and estimated periods for 23 RR Lyrae stars.
The follow-up study by van Agt \& Oosterhoff (1959) presents 37 variables and
periods were reported for 35 variables.
Sawyer Hogg (1973) catalogue presents 42 variables in the cluster including van Agt \& Oosterhoff (1959) 
and Terzan et al. (1973) variables.
Clement (1990) and Clement et al. (1993) studied the cluster to revise the periods and to investigate for the double mode
RR Lyrae pulsators. 
All the above work was based on photographic plates. 
Brocato et al. (1994) derived periods of the variables
using CCD observations. An extensive BVI CCD photometric study by Walker (1994) added new variables
and also found new double mode RR Lyrae and SX Phe stars.

In the light of above discussion, we also monitored the cluster NGC 4590 to search for new variables and to 
check if the known variables show any significant change in the periods.  
We detected 40 known variables in the present analysis.
Comparison of the present periods with Walker (1994) exhibit no significant change in the period.
The deviation in the periods is $\sim$ 0.0001--0.001d. For V26 and V35, our periods are in agreement with 
Rosino \& Pietra (1954) and van Agt \& Oosterhoff (1959), respectively.
Blazhko effect has been found in RRab type stars and comparison with Walker (1994)
shows a significant change in the Blazhko effect for V25 and little variation for few other stars.
A careful inspection of the light curves provides 9 probable new variables.
The equatorial coordinates, periods and phased light curves for 
49 (40 known + 9 new discovered) variables have been 
provided in the present study. 

The structure of the article is as follows: 
Observations, reduction procedures and astrometry are described in Sect.~\ref{OBS}.
Section~\ref{per} deals with the method for variables identification and estimating periods.
Variability characteristics of the variables found in the present study are discussed in Sect.~\ref{char}.
In Sect.~\ref{member}, we provide the membership of few variables.
We conclude our work in Sect.~\ref{con}. 
%__________________________________________________________________

\section{Observations and data reduction}
\label{OBS}

The observations of NGC 4590 were carried out in Johnson $V$ filter using
104-cm Sampurnanand telescope ($f$/13). The telescope is located at ARIES, Manora Peak, 
Nainital, India. The cluster was observed for 10 nights during January--March, 2011 
using a 2048$\times$2048 CCD in 2$\times$2 binning mode, providing a scale of ~0.73 arcsec pixel$^{-1}$.
The complete log of observations is given in Table~\ref{log}.

In order to perform the CCD image processing, we used the standard procedure of bias subtraction, flat-fielding 
and cosmic rays removal with IRAF\footnote{IRAF is distributed by the National Optical Astronomical
Observatory which is operated by the Association of Universities for Research in Astronomy, 
under contact with the National Science Foundation}. The instrumental magnitudes of the stars were 
computed using the DAOPHOT package (Stetson 1987). We performed both aperture and point spread function (PSF)
photometry to get better results in the crowded regions towards the cluster center. In order to find the translation,
rotation and scaling solutions between different photometry files, we used DAOMATCH (Stetson 1992) package. In
addition to this, we have to correct for frame-to-frame flux variations also due to varying airmass. 
In our data, the exposure time is uniform (180 s) for all frames.
We used DAOMASTER (Stetson 1992) to perform this correction. This task uses an additive constant to 
make the mean flux level of each frame equals to the reference frame. 
The reference frame chosen is the first image observed on 06 March 2011, 
as this night have maximum number of frames.
From the output of DAOMASTER, 
we get corrected magnitudes listed
in a ``.cor'' file. This file is further used to search for variable stars.

%%%%%%%%%%%%%%%%%%%%%%%%%%%%%%%%%%%%%%%%%%%%%%%%%%%%%%%%%%%%%%%%%%%%%%%%%%%%%%%%%%%%%%%%%
\subsection{Astrometry}

The next step is to transfer CCD pixel coordinates of the stars to the celestial 
coordinates in J2000.0 equinox. For this purpose, we used the online digitized sky 
ESO catalogue in SkyCat software as an absolute astrometric reference frame.
We have used IRAF tasks CCMAP and CCTRAN to carry out the transformations. 
The transformations have an rms value of about 0.005 and  0.7 arcsec
in right ascension (RA) and declination (DEC) respectively.
  
\section{Variable identification and periods determination} 
\label{per}

\begin{figure*}
\hspace{-2.5cm}
\includegraphics{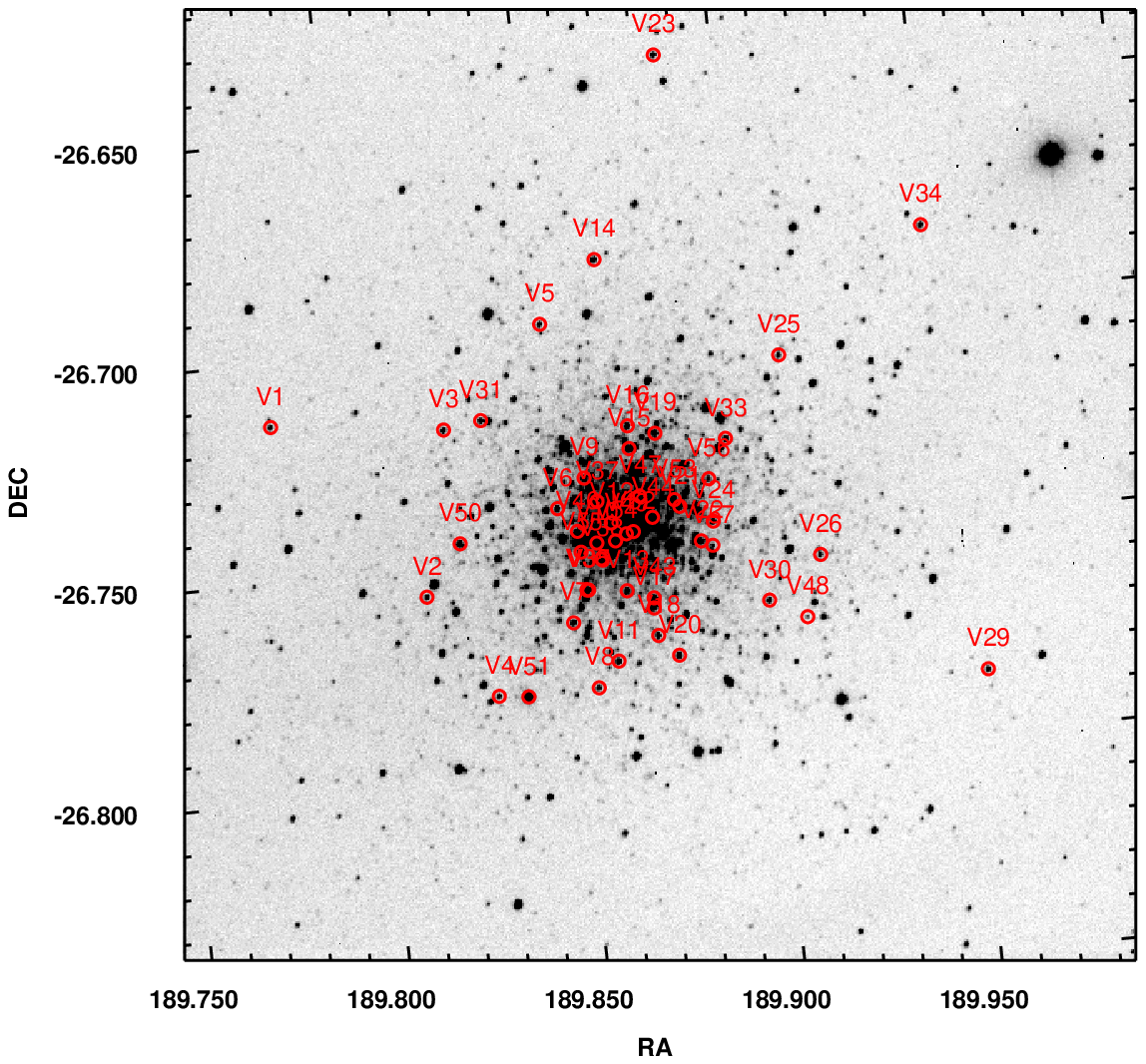}
\caption{ The observed region of NGC 4590 taken with
104 cm Sampurnanand telescope in $V$ band. The open circles represent the variables
identified in the present work. RA and DEC are given in degree.}
\label{idvar}
\end{figure*}

The differential instrumental magnitudes vs Julian dates (JD) 
for the stars, were plotted from ``.cor'' file.
Variable stars were selected using the classical light curve viewing method.  
We performed differential photometry, which is an important tool
for time series study. It has the advantage of being effective even during thin cloudy nights,
and program stars and comparison stars having the same airmass and exposure time.
For a program star, 3 comparison stars ( C1, C2, C3 ) are chosen
in such a way that they have almost same instrumental 
magnitude as the program star. The difference of instrumental magnitudes for C1, C2 and C3 i.e., 
(C1-C2), (C1-C3) and (C2-C3) remain almost constant with time.
Figure~\ref{idvar} shows the reference image in which we also plotted the variables with open circles.
For calculating the periods, we used Lomb-Scargle(LS) periodogram (Lomb 1976; Scargle 1982). This method
works well with unevenly sampled data (Lata et al. 2011, Lata et al. 2012). We calculated periods with the 
software period main available at starlink\footnote{http://www.starlink.uk} using the LS algorithm. 
We further confirmed the periods using NASA exoplanet archive periodogram 
service\footnote{http://exoplanetarchive.ipac.caltech.edu/cgi-bin/Periodogram/nph-simpleupload}.
For the stars showing spurious periods, we visually inspected the phased light curves and opted
the period showing the best light curve.
Conclusively, we present phased light curves for the known and new probable variable candidates,
where the $\Delta$$V$ is ``differential instrumental magnitude''

\section{Characteristics of variable stars}
\label{char}

\subsection{Known variables}

\begin{figure}
\centering
\includegraphics[width=12cm]{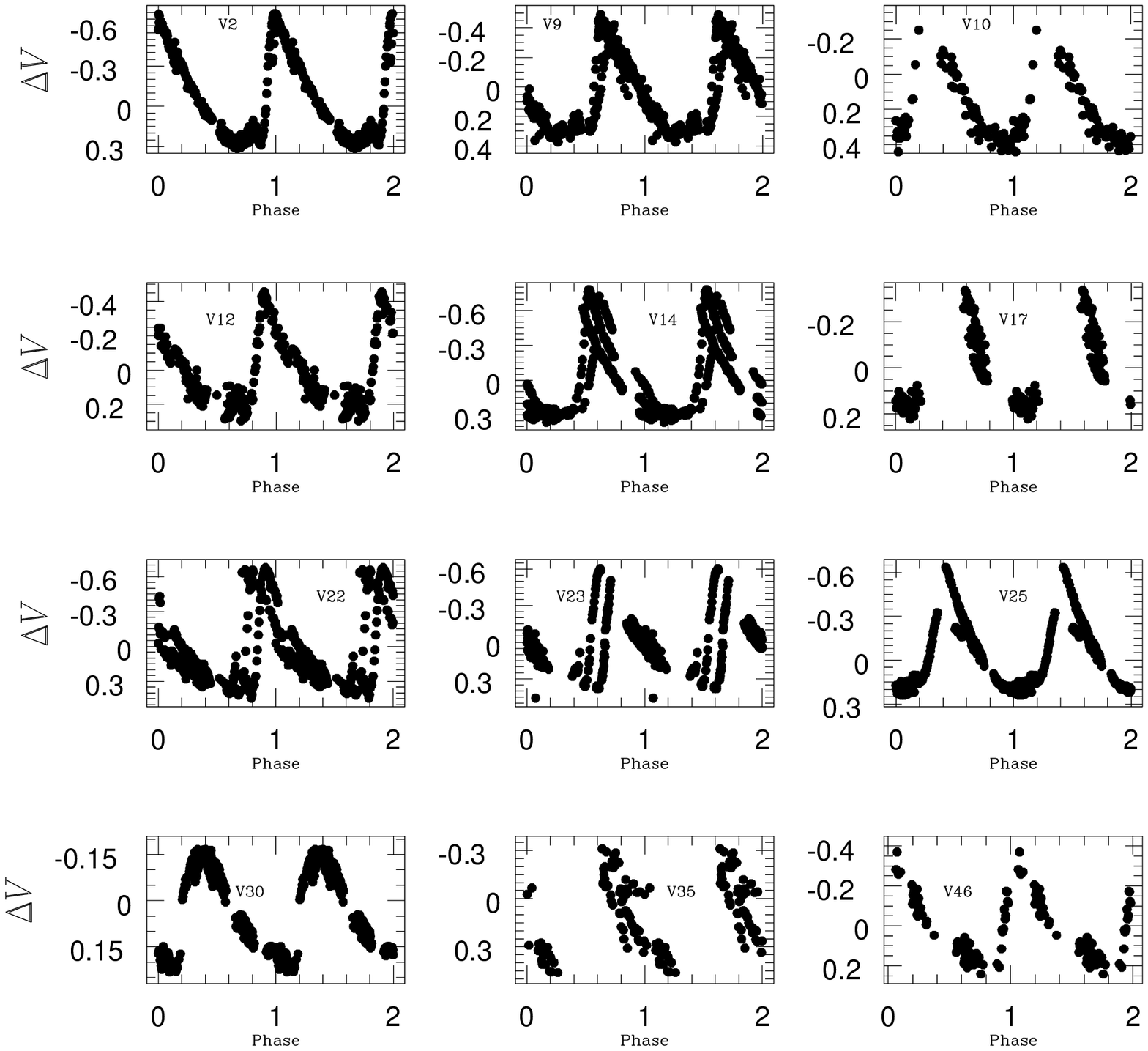}
\caption{ Phased light curves for known RRab type stars. The $\Delta$$V$ shows
differential instrumental magnitude .}
\label{RRab}
\end{figure}

\begin{figure}
\centering
\includegraphics[width=12cm]{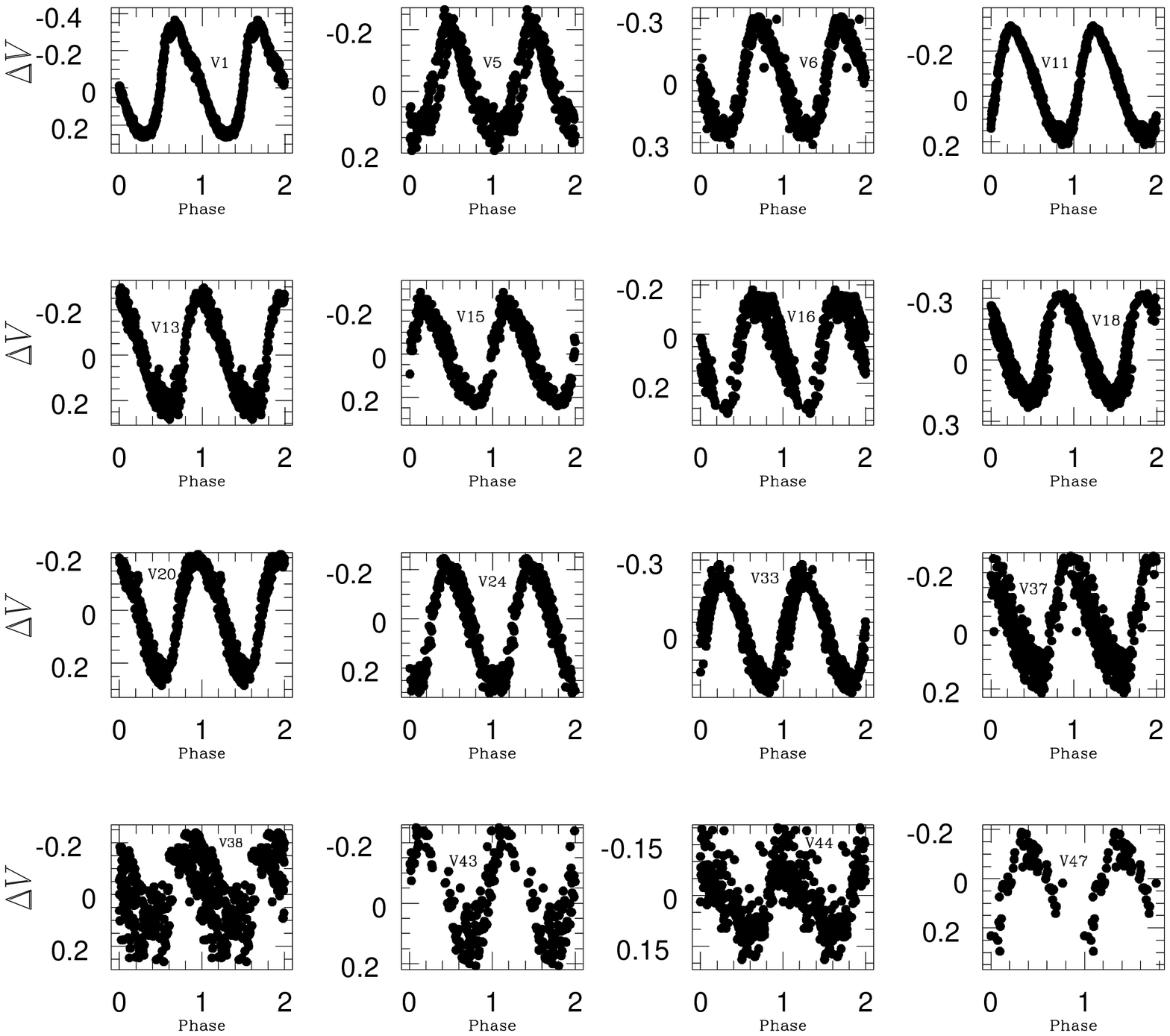}
\caption{ Phased light curves for known RRc type stars. The $\Delta$$V$ shows
differential instrumental magnitude.}
\label{RRc}
\end{figure}

\begin{figure}
\centering
\includegraphics[width=12cm]{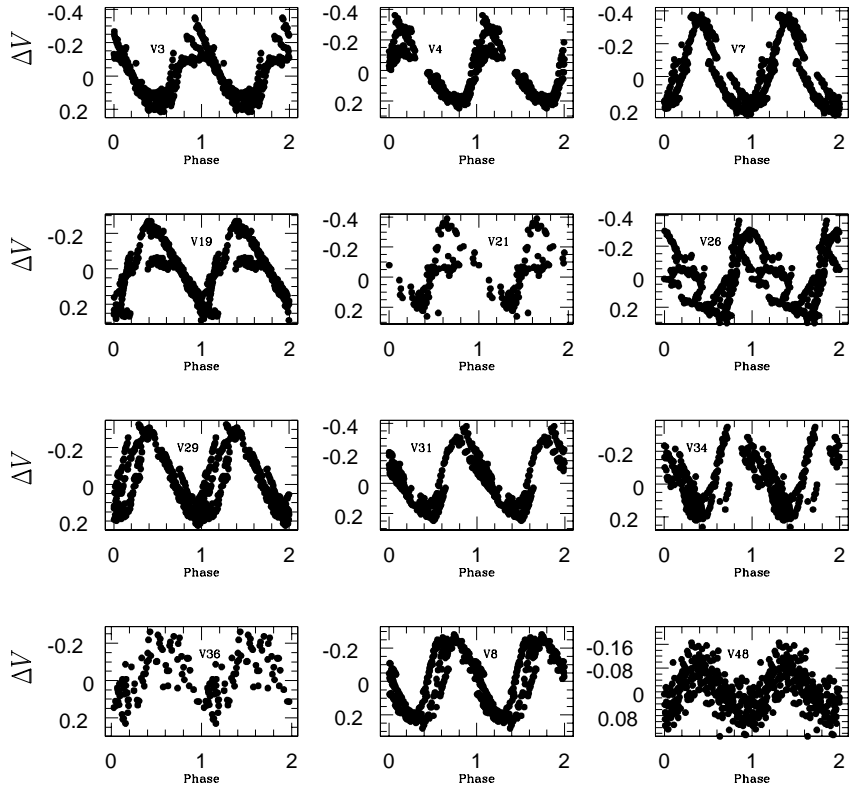}
\caption{ Phased light curves for known RRd stars (double-mode pulsators). V48 is a SX Phe star.}
\label{RRd}
\end{figure}

NGC 4590 has 48 known variable stars, and among them we found 40 variables
in the present analysis.
We could not find V27, V28, and V32 as they were out of our observational area.
Few known variables could not be detected because of their location in the crowded region of the cluster.
Table~\ref{varold} lists the equatorial coordinates, variable type and period of the detected variables.
The phased light curves of the variables are shown in Fig~\ref{RRab}, \ref{RRc} and \ref{RRd}
for RRab, RRc and RRd type variables, respectively. One SX Phe star V48 is also plotted in  Fig~\ref{RRd}.

The periods alongwith the other parameters of known variables are listed in Table~\ref{varold}. Column 1 
lists the variable ID
as listed in the literature, while Cols. 2 and 3 give the equatorial coordinates in J2000.0. 
Finally, the classification of variables 
and periods derived in the present study are given in Cols. 4 and 5 respectively. Our derived periods match well 
with the Walker (1994) values.
For V26, present value of period (0.413578d) matches more closely with 
Rosino and Pietra (1954) period (0.413217d), while
for V35, our period (0.719225d) is closer to van Agt \& Oosterhoff (1959) period (0.71608d).
V35 is located in crowded region of the cluster, while V26 is in relatively sparse region.
We detected 12 RRab type and 16 RRc Bailey type stars.
Clement (1990) and Clement et al. (1993) investigated the cluster to find double mode pulsators and listed 
fundamental mode and first overtone periods of 8 RRd variables, and discussed about V21 being a promising RRd candidate. 
Walker (1994) found the fundamental mode and first overtone periods for V21 and also 
added V8, V36 and V45 to the double mode pulsators.
We found 11 RRd stars and listed their fundamental mode period in Table~\ref{varold}. 
The present period values for all RRd stars match with 
the fundamental periods given by Walker (1994).

RRab stars are fundamental mode pulsators 
showing a fast rise and a slower decline in their light curves.
Figure~\ref{RRab} shows the phased light curves for RRab stars. 
V10 is lying near the cluster center 
in the crowded region, and therefore, it could not be detected in all the images. 
V17 and V35 are very close to V43 and V36, respectively and hence, they have
more scattering in their light curves.

Some RRab stars of NGC 4590 show a pronounced Blazhko effect.
Blazhko effect, discovered by Blazhko (1907), is the modulation in the light curve variations
of the RR Lyrae stars. 
A review by Stellingwerf (1976) provided six mechanisms as possible reason for the occurrence of
the Blazhko effect. 
Kolenberg et al. (2006) and Kolenberg (2011) discussed the new insights about the peculiarities of Blazhko stars.
According to Kolenberg et al. (2006) and Kolenberg (2011), two main mechanisms for the occurrence of Blazhko effect 
are: the resonance model (e.g., Dziembowski \& Mizerski 2004, and references therein)
and the magnetic model (see, e.g., Shibahashi 2000). 
Both of these models give the excitation of nonradial pulsation mode components 
in the modulated star as the reason for Blazhko effect.
According to the resonance model, excitation of nonradial dipole mode (degree $l$ = 1)
takes place, while magnetic models believe that the magnetic field deforms the radial mode
to have additional quadrupole components (degree $l$ = 2).
Jurcsik et al. (2002, 2005) found that the Blazhko stars are ``always distorted'', even at the maximum 
amplitude and the possible maximum value of the modulation frequency depends on the pulsation frequency.
Sothers (2006) explained the Blazhko effect as the turbulent convection of the hydrogen and helium ionization zones
becoming cyclically weakened and strengthened which is caused by the presence of a transient magnetic field generated by
either a turbulent or a rotational dynamo mechanism. Most recently, Gillet (2013) stated that 
the Blazhko stars are located in the region of the instability strip where fundamental and first overtone modes are excited
at the same time. Gillet (2013) gave occurrence of shocks as the reason behind the Blazhko effect.
Skarka \& Zejda (2013) presented RRc type stars which show changes in Blazhko effect.
For some RR Lyrae stars, Blazhko effect nearly vanishes after some years (Smith, 1995).

Blazhko effect is seen for RRab stars V14, V22 and V23, as shown in Fig.~\ref{RRab}.
Walker (1994) found Blazhko effect in V2 and V9. For both the stars,
it is very less prominent in our work as compared to Walker (1994).
As documented by Walker (1994), stars like V14 exhibit amplitude shift and possibly phase shifts
of the ``bump'' with respect to the maximum. 
Therefore, it requires very precise photometry for their detection and better light curves.
Most noticeable point in the light curve of V25
is that it shows significant Blazhko effect in Walker (1994) study, having cycle-cycle
variation, while present work shows no such signature.

RRc stars are RR Lyrae stars with nearly sinusoidal light curves pulsating in the first overtone mode.
Figure~\ref{RRc} shows light curves for RRc type stars. 
V5 has the shortest period among 
RRc variables and this was also found by Walker (1994). There is significant scattering for V38 and V44, 
because they lie in the central region of the cluster.
V43 is blended with one or more stars, that include V17. V47 has less data points too, due to crowding.
Clement et al. (1993) suggested V33 to be RRc type variable. 
They stated that it could be a double mode pulsator in 1950, but the pulsation properties of V33 changed dramatically 
between 1950 and 1986. It has been almost 19 years since last variable stars studies were carried out for the cluster.
So, it is interesting to see if V33 has become a RRd star again. 
Inspection of its light curve gives no clue of double mode pulsations
and it remains a RRc star.

Double mode RR Lyrae pulsators (RRd) are the stars which pulsate in both 
fundamental and first overtone mode.
The periods for these stars lie in the interval 0.390--0.429d (Smith 1995).
The present periods for RRd stars are more than
0.39d, except for V7, for which it is 0.38833d. 
Light curves for reported double mode pulsators are given in Fig~\ref{RRd}. 
The detection of star V21 is difficult because of crowding.
The same was suggested by Clement et al. (1993).
We have less number of data points and more scattering for this star.
V35 is in the crowded region and V36 is very near to it. 
This might be the cause of scattering
for this star.

V48 is a SX Phe star, for which Walker (1994) found period as 1.0375 h (=0.043229167d).
We calculated its period as 0.043217 d. Its light curve is also plotted in Fig~\ref{RRd}.
Light curve of V48 has more scattering, due to faintness of SX Phe stars.
 
\subsection{New probable variables}

\begin{figure}
\centering
\includegraphics[width=12cm]{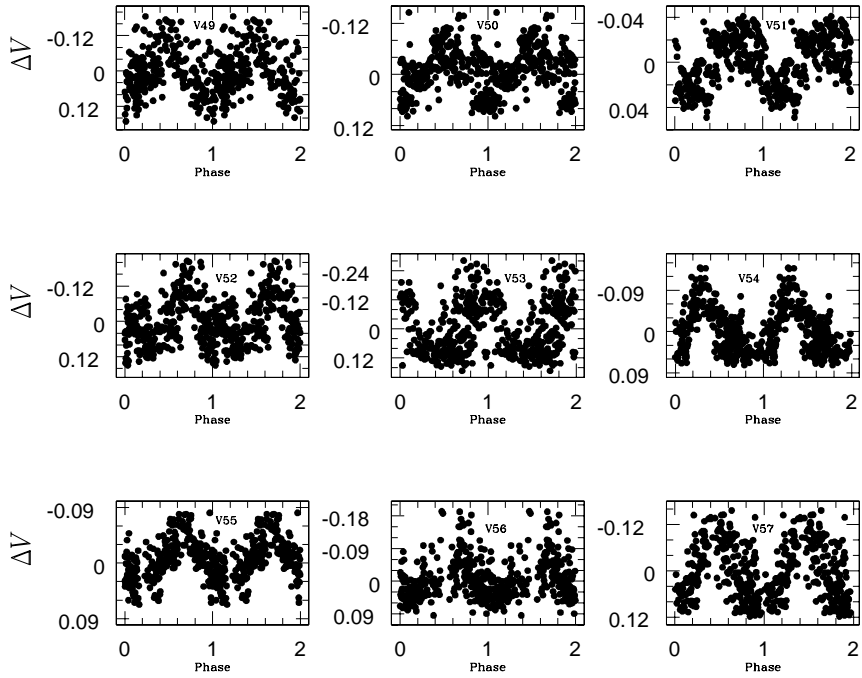}
\caption{ Phased light curves, differential instrumental magnitude ($\Delta$$V$) vs phase for probable new variable stars.}
\label{new}
\end{figure}

On careful inspection of the light curves of more stars in our data, we found some
probable variable candidates. In Table~\ref{varnew}, we list their equatorial coordinates
and periods. Their phased light curves have been plotted in Fig.~\ref{new}. 
Due to the location in the crowded region, they show scattering in their light curves.
It can be noticed that for V49--V53, period lies in the range 0.33-0.43d. 
Their light curves show less amplitude. For V49, V50 and V52 
the amplitude is $\sim$0.30 mag, while 
V51 has even smaller amplitude ($\sim$0.10 mag).
Photometric errors for V51 star are better than 0.01 mag.
Amplitude for V53 is $\sim$0.45 mag.
Based on their period and shape of the light curves, V49--V53 can be classified as 
RRc type variables.

The periods for stars V54 to V57 are in the range 0.51-0.54d, suggesting them to be RRab type stars.
But the shape of the light curves of these variables do not seem to match with the shape of RRab type
variables. The amplitudes of these variables range from $\sim$0.20 to 0.30 mag. 
Based on amplitude, they could be RRc type stars, but the period values do not agree with this fact. 
Thus, because of these two contradictory aspects, 
it is very difficult to classify these objects based on their variability characteristics.
In order to confirm the nature of these 
variables, better resolution data in the cluster core is highly required.

\section{Membership of variables}
\label{member}
Lane et al. (2011) conducted spectroscopic studies of 10 globular clusters,
including NGC 4590 using AAOmega instrument on the Anglo-Australian Telescope (AAT).
They estimated heliocentric radial velocity and stated the status of membership
for 990 stars in the cluster region.
Membership criteria has been discussed in detail by Lane et al. (2011) and references therein.
We matched our catalogue with Lane et al. (2011) data
within 2 arcsec match-up radius and found 4 common stars which are given in Table~\ref{lane}.
Columns 1 and 2 give variable ID and the Lane et al. (2011) ID respectively.
Columns 3 and 4 show the heliocentric radial velocity and corresponding error in Km/s, 
while Col. 5 affirms the membership status as positive for all 4 variables.
The heliocentric velocity for NGC 4590 is listed as -94.3$\pm$0.4 Km/s by Harris (1996).
Radial velocities for all 4 stars in Table~\ref{lane} are in agreement with
Harris (1996) value within 3$\sigma$ errors.
Hence, V10, V21, V50 and V51 are likely cluster members.

\section{Conclusions}
\label{con}

We have carried out time series photometry of the globular cluster NGC 4590 
using images in $V$ band for 10 nights.
The results of the present study can be summarized in the following way:

\begin{enumerate}

\item Phased light curves and periods for 40 known variables
have been revised. We did not find any significant change in the periods or in the variability type.

\item RRab stars show some change in Blazhko effect. 
Variable V25 has distinguishable change since last study by Walker (1994).

\item We find 9 new probable variables. We present their phased light curves and periods. 
Five new probable variables are categorized as Bailey type RRc stars, while for 4 variables
only periods and phased light curves are provided. The classification 
for these 4 variables remains undetermined.

\item The status of membership of variable stars has been assigned on the basis of comparison with
Lane et al. (2011). Variable stars V10, V21, V50, V51 are found to be cluster members.
\end{enumerate}

\section{ACKNOWLEDGEMENTS}
Authors are thankful to anonymous referee for constructive comments.
D.P. Sariya acknowledges Prof. Ram Sagar for motivation to the investigation work of 
variable stars in globular clusters.

%\bibliographystyle{model1a-num-names}

%%%%%%%%%% Figures are here
%%%%%%%%%%%%%%%%%%%%%%%%%%%%%%%%%%%%%%%%%%%%%%%%%%%%%%%%%%%%%%%%%%%%%%%%%%%%%%%%%%%%%%%% 
\clearpage
%%%%%%%%%%%%%%%%%%%%%%%%%%%%%%%%%%%%%%%%%%%%%%%%%%%%%%%%%%%%%%%%%%%%%%%%%%%%%%%%%%%%%%%%%
%%%%%%%%%%%%%%%%%%%%%%%%%%%%%%%%%%%%%%%%%%%%%%%%%%%%%%%%%%%%%%%%%%%%%%%%%%%%%%%%%%%%%%%%%

\clearpage
%%%%%%%%%%%%%%%%%%%%%%%%%%%%%%%%%%%%%%%%%%%%%%%%%%%%%%%%%%%%%%%%%%%%%%%%%%%%%%%%%%%%%%%%%

\clearpage
%%%%%%%%%%%%%%%%%%%%%%%%%%%%%%%%%%%%%%%%%%%%%%%%%%%%%%%%%%%%%%%%%%%%%%%%%%%%%%%%%%%%%%%%%

\clearpage
%%%%%%%%%%%%%%%%%%%%%%%%%%%%%%%%%%%%%%%%%%%%%%%%%%%%%%%%%%%%%%%%%%%%%%%%%%%%%%%%%%%%%%%%%

\clearpage
%%%%%%%%%%%%%%%%%%%%%%%%%%%%%%%%%%%%%%%%%%%%%%%%%%%%%%%%%%%%%%%%%%%%%%%%%%%%%%%%%%%%%%%%

%%%%%%%%%%%%%%%%%%%%%%%%%%%%%%%%%%%%%%%%%%%%%%%%%%%%%%%%%%%%%%%%%%%%%%%%%%%%%%%%%%%%%%%% are here

%%%%%%%%%%%%%%%%%%%%%%%%%%%%%%%%%%%%%%%%%%%%%%%%%%%%%%%%%%%%%%%%%%%%%%%%%%%%%%%%%%%%%%%%%%%%%%%%

\clearpage
\begin{table}
\caption{Fundamental parameters of NGC~4590 taken from Harris (1996) (2010 edition).
}
\centering
\label{par}
\begin{tabular}{cc}
\\
\hline\hline Parameters & Values \\ \hline
$\alpha$(J2000)  &     12$^{\rm h}$ 39$^{\rm m}$ 27$^{\rm s}$.98  \\
$\delta$(J2000)  & $-26^\circ$ 44$'$ 38.6$''$    \\
$l$              & $299^\circ.63$                \\
$b$              & $36^\circ.05$              \\
$\rm [Fe/H]$     & $-2.23 $           \\
$E(B-V)  $       & 0.05 mag \\
$(m-M)_V$          & 15.21 mag\\
\hline
\end{tabular}
\end{table}

%%%%%%%%%%%%%%%%%%%%%%%%%%%%%%%%%%%%%%%%%%%%%%%%%%%%%%%%%%%%%%%%%%%%%%%%%%%%%%%%%%%  
%%%%%%%%%%%%%%%%%%%%%%%%%%%%%%%%%%%%%%%%%%%%%%%%%%%%%%%%%%%%%%%%%%%%%%%%%%%%%%%%%%%%%%%%%%

\clearpage
\begin{table}
\caption{ Description of our observations. All the images are taken in $V$ filter.}
\centering
\label{log}
\begin{tabular}{ccc}
\\
\\
\hline\hline S. No. & Date of observations & No. of frames \\
\hline
1    &    06 Jan 2011     &   14 \\
2    &    07 Jan 2011     &   13\\
3    &    08 Jan 2011     &   18\\
4    &    10 Jan 2011     &   28\\
5    &    01 Feb 2011     &   50\\
6    &    02 Feb 2011     &   31\\
7    &    03 Feb 2011     &   30\\
8    &    06 Mar 2011     &   50\\
9    &    08 Mar 2011     &   27\\
10   &    09 Mar 2011     &   26\\
\hline
\end{tabular}
\end{table}
%%%%%%%%%%%%%%%%%%%%%%%%%%%%%%%%%%%%%%%%%%%%%%%%%%%%%%%%%%%%%%%%%%%%%%%

%%%%%%%%%%%%%%%%%%%%%%%%%%%%%%%%%%%%%%%%%%%%%%%%%%%%%%%%%%%%%%%%%%%%%%%%%%%%%%%%%%%%%%%%%

\clearpage
\begin{table}
\caption{Parameters for the known variables found in the present study.
The first column denotes the variable ID listed in the literature. Period values (in days) given by Walker (1994)
are also listed in the last column.}
\centering
\label{varold}
\begin{tabular}{cccccc}
\\
\hline\hline Var ID & RA & DEC&Variable & Period & Period (d) by  \\
 & (2000.0) & (2000.0) &  Type & (days) & Walker (1994)\\
\hline

  V1  &   12:39:06.96   &  -26:42:52.0   &  RRc   & 0.349503 & 0.34960\\
  V2  &   12:39:15.37   &  -26:45:23.6   &  RRab  & 0.578402 & 0.57816\\
  V3  &   12:39:17.44   &  -26:43:08.4   &  RRd   & 0.390790 & 0.39078\\
  V4  &   12:39:19.13   &  -26:46:50.6   &  RRd   & 0.395769 & 0.39602\\
  V5  &   12:39:23.94   &  -26:41:50.0   &  RRc   & 0.282306 & 0.28211\\
  V6  &   12:39:23.86   &  -26:44:22.1   &  RRc   & 0.368656 & 0.36859\\
  V7  &   12:39:24.11   &  -26:45:56.7   &  RRd   & 0.388336 & 0.38803\\
  V8  &   12:39:25.25   &  -26:46:51.9   &  RRd   & 0.390790 & 0.39044\\
  V9  &   12:39:25.66   &  -26:43:59.3   &  RRab  & 0.578402 & 0.57904\\
  V10 &   12:39:26.02   &  -26:44:53.4   &  RRab  & 0.551107 & 0.55127\\
  V11 &   12:39:26.62   &  -26:46:31.7   &  RRc   & 0.364922 & 0.36494\\
  V12 &   12:39:27.02   &  -26:44:38.2   &  RRab  & 0.615614 & 0.61580\\
  V13 &   12:39:27.56   &  -26:45:34.9   &  RRc   & 0.361730 & 0.36176\\
  V14 &   12:39:27.65   &  -26:41:01.6   &  RRab  & 0.557850 & 0.55680\\
  V15 &   12:39:28.57   &  -26:43:38.9   &  RRc   & 0.372468 & 0.37227\\
  V16 &   12:39:28.62   &  -26:43:20.2   &  RRc   & 0.381644 & 0.38201\\
  V17 &   12:39:29.10   &  -26:45:51.2   &  RRab  & 0.666618 & 0.66755\\
  V18 &   12:39:29.19   &  -26:46:14.0   &  RRc   & 0.367057 & 0.36735\\
  V19 &   12:39:30.24   &  -26:43:28.5   &  RRd   & 0.396401 & 0.39170\\
  V20 &   12:39:30.33   &  -26:46:31.9   &  RRc   & 0.385542 & 0.38577\\
  V21 &   12:39:31.29   &  -26:44:30.3   &  RRd   & 0.401513 & 0.40707\\
  V22 &   12:39:32.38   &  -26:45:00.3   &  RRab  & 0.562395 & 0.56344\\
  V23 &   12:39:32.54   &  -26:38:19.3   &  RRab  & 0.659716 & 0.65892\\
  V24 &   12:39:33.23   &  -26:44:45.1   &  RRc   & 0.376551 & 0.37647\\
  V25 &   12:39:38.25   &  -26:42:34.7   &  RRab  & 0.641207 & 0.64149\\
  V26 &   12:39:39.54   &  -26:45:21.2   &  RRd   & 0.413578 & 0.40696\\
  V29 &   12:39:49.00   &  -26:47:08.4   &  RRd   & 0.394770 & 0.39522\\
  V30 &   12:39:36.15   &  -26:45:54.3   &  RRab  & 0.734163 & 0.73364\\
  V31 &   12:39:19.75   &  -26:43:03.8   &  RRd   & 0.399592 & 0.39965\\
  V33 &   12:39:34.49   &  -26:43:38.5   &  RRc   & 0.390782 & 0.39059\\
  V34 &   12:39:47.68   &  -26:41:00.0   &  RRd   & 0.400504 & 0.40010\\
  V35 &   12:39:25.23   &  -26:45:31.1   &  RRab  & 0.719225 & 0.70250\\
  V36 &   12:39:25.13   &  -26:45:30.9   &  RRd   & 0.426776 & 0.41099\\
  V37 &   12:39:26.29   &  -26:44:19.4   &  RRc   & 0.384700 & 0.38463\\
\end{tabular}
\end{table}
%%%%%%%%%%%%%%%%%%%%%%%%%%%%%%%%%%%%%%%%%%%%%%%%%%%%%%%%%%%%%%%%%%%%%%%%%%%%%

\clearpage
\begin{table}
\centering
\begin{tabular}{cccccc}
  V38 &   12:39:26.20   &  -26:45:06.8   &  RRc   & 0.388116 & 0.38294\\
  V43 &   12:39:29.15   &  -26:45:42.9   &  RRc   & 0.370644 & 0.37071\\
  V44 &   12:39:29.56   &  -26:44:37.0   &  RRc   & 0.385321 & 0.38514\\
  V46 &   12:39:24.92   &  -26:44:42.3   &  RRab  & 0.739256 & 0.73816\\
  V47 &   12:39:28.91   &  -26:44:18.5   &  RRc   & 0.372757 & 0.37295\\
  V48 &   12:39:38.36   &  -26:46:11.0   &  SX Phe& 0.043217 & 0.04323\\
\hline
\end{tabular}
\end{table}
%%%%%%%%%%%%%%%%%%%%%%%%%%%%%%%%%%%%%%%%%%%%%%%%%%%%%%%%%%%%%%%%%%%%%%%%%%%%%%%%%%%%%%%%%
%%%%%%%%%%%%%%%%%%%%%%%%%%%%%%%%%%%%%%%%%%%%%%%%%%%%%%%%%%%%%%%%%%%%%%%%%%%%%%%%%%%%%%%%%

\clearpage
\begin{table}
\caption{Light curve parameters for probable new variables. ID given by us,
is in continuation to the list of known variables.}
\centering
\label{varnew}
\begin{tabular}{ccccc}
\\
\hline\hline Var ID & RA(2000.0)& DEC(2000.0)&Variable Type & Period(days)\\
\hline
  V49 &   12:39:27.87 &  -26:44:48.3  & RRc & 0.370135 \\
  V50 &   12:39:17.72 &  -26:44:42.7  & RRc & 0.334568 \\
  V51 &   12:39:20.92 &  -26:46:53.5  & RRc & 0.423975 \\
  V52 &   12:39:28.31 &  -26:44:47.1  & RRc & 0.349049 \\
  V53 &   12:39:31.00 &  -26:44:24.3  & RRc & 0.269895 \\
  V54 &   12:39:27.17 &  -26:44:53.2  & ?   & 0.536981 \\
  V55 &   12:39:25.01 &  -26:44:59.6  & ?   & 0.535583 \\
  V56 &   12:39:33.22 &  -26:44:10.1  & ?   & 0.511493 \\
  V57 &   12:39:33.05 &  -26:45:04.8  & ?   & 0.535683 \\
\hline
\end{tabular}
\end{table}
%%%%%%%%%%%%%%%%%%%%%%%%%%%%%%%%%%%%%%%%%%%%%%%%%%%%%%%%%%%%%%%%%%%%%%%%%%%%%%%%%%%%%%%%%

\clearpage
\begin{table}
\caption{Membership status of variables as a result of comparison with Lane et al. (2011) catalogue.
Heliocentric radial velocities ($V_R$) and their errors ($eV_R$) taken from Lane et al. (2011), are listed in columns 3 and 4.
Last column shows the membership status of variables.}
\centering
\label{lane}
\begin{tabular}{ccccc}
\\
\hline\hline Var ID & Lane ID & $V_R$ & $eV_R$ & Membership\\
 & & (Km/s)&(Km/s)&\\
\hline
  V10 & 215   & -99.7   & 1.5  & yes \\
  V21 & 330   & -94.6   & 1.6  & yes \\
  V50 & 811   & -96.9   & 1.9  & yes \\
  V51 &  58   & -95.4   & 1.5  & yes \\
\hline
\end{tabular}
\end{table}
%%%%%%%%%%%%%%%%%%%%%%%%%%%%%%%%%%%%%%%%%%%%%%%%%%%%%%%%%%%%%%%%%%%%%%%%%%%%%%%%

\end{document}